# Solitons in non-paraxial optics

D. DAKOVA[b], A. DAKOVA[a,b], V. SLAVCHEV[a,c*], L. KOVACHEV[a]
[a]Institute of Electronics, Bulgarian Academy of Sciences, 72 Tzarigradsko shossee,1784 Sofia, Bulgaria
[b]Faculty of Physics, University of Plovdiv "Paisii Hilendarski", 24 Tsar Asen Str., 4000 Plovdiv, Bulgaria
[c]Faculty of Pharmacy, Medical University - Plovdiv, Bul. Vasil Aprilov 15-A, 4002 Plovdiv, Bulgaria

The well-known (1+1D) nonlinear Schrödinger equation (NSE) governs the propagation of narrow-band pulses in optical fibers and others one-dimensional structures. For exploration the evolution of broad-band optical pulses (femtosecond and attosecond) it is necessary to use the more general nonlinear amplitude equation (GNAE) which differs from NSE with two additional non-paraxial terms. That is way, it is important to make clear the difference between the solutions of these two equations.
We found a new analytical soliton solution of GNAE and compare it with the well-known NSE one. It is shown that for the fundamental soliton the main difference between the two solutions is in their phases. It appears that, this changes significantly the evolution of optical pulses in multisoliton regime of propagation and admits a behavior different from that of the higher-order NSE solitons.

## 1. Introduction

In recent years actively are studied the phenomena resulting from the evolution of ultra-short optical pulses in dispersive nonlinear medium [1,2,3]. In the femtosecond region it is quite easy to be obtained broad-band phase-modulated pulses or to reach the region of *5-15 fs* where $\Delta\omega \approx \omega_0$. One dimensional NSE is derived for narrow-band pulses ($\Delta\omega << \omega_0$) propagating in single-mode fibers or planar waveguides [4-7]. That is way, it is important to study the behavior of laser pulses with broad-band spectrum in such structures.

And here appears the basic question: What kind of equation describes the propagation of broad-band pulses as well as narrow-ones? To answer this question we turned to the well-known nonlinear amplitude equation derived in [8] where in addition to the dispersion extra non-paraxial terms exist. The dimensionless analysis performed below shows that these terms play a significant role in the dynamics of light pulses.

Our quick review shows that, up to now, the dynamics of laser solitons in different nonlinear fibers and optical materials, based on the use of NSE, modified by adding extra terms, is well studied [9-18]. However, exact analytical soliton solutions of GNAE are not found. This is way one of the main tasks of this paper is to find such solution, highlighting its advantages and in addition to make a comparison with NSE-one. The NSE and GNAE are written in local time coordinate system. In our investigations are neglected the losses and the higher order nonlinear and dispersive effects.

## 2. Exact analytical solution of the nonlinear amplitude equation

### 2.1 Basic equation

The equation governing the evolution of broad-band and narrow-band pulses in Kerr-type nonlinear dispersive isotropic medium, is [2,8,19]:

$$2ik_0\left[\frac{\partial A}{\partial z}+\frac{1}{v_{gr}}\frac{\partial A}{\partial t}\right]-k_0k''\frac{\partial^2 A}{\partial t^2}+\frac{\partial^2 A}{\partial z^2}-\frac{1}{v_{gr}^2}\frac{\partial^2 A}{\partial t^2}+k_0^2 n_2|A|^2 A = 0 \quad (1)$$

The GNAE is obtained for the first time in [8] and if the fourth and the fifth non-paraxial terms in equation (1) are neglected it can be transformed into the standard NSE. It is important to be pointed, that equation (1) is obtained after using Taylor series of the $k_2(\omega)$ near to main frequency $\omega_0$. This series continue to be strongly cognate up to single cycle regime. The equation (1) governs correctly the evolution of narrow-band pulses as well as pulses with *5-6 fs* duration.

By making the substitution below, equation (1) can be presented in local time coordinate system $T = t - \frac{z}{v_{gr}}, z=z$:

$$2ik_0\frac{\partial A}{\partial z}+\frac{\partial^2 A}{\partial z^2}-\frac{2}{v_{gr}}\frac{\partial^2 A}{\partial z \partial T}-k_0 k''\frac{\partial^2 A}{\partial T^2}+k_0^2 n_2|A|^2 A = 0 \quad (2)$$

The main difference between GNAE and NSE is in the additional two terms – second derivative ($\partial^2 A/\partial z^2$) and mixed derivative ($\partial^2 A/\partial z \partial T$). To perform a quantitative analysis of the influence of the different terms in equation (2) we transform it in dimensionless form by changing the variables:

$$\xi = \frac{z}{z_0}, \tau = \frac{T}{T_0}, z_0 = v_{gr}T_0, A(\tau,\xi) = A_0 U(\tau,\xi) \quad (3)$$

$$i\frac{\partial U}{\partial \xi}+\frac{1}{2\alpha}\left(\frac{\partial^2 U}{\partial \xi^2}-2\frac{\partial^2 U}{\partial \xi \partial \tau}\right)+\frac{|\beta|}{2\alpha}\frac{\partial^2 U}{\partial \tau^2}+\gamma |U|^2 U = 0 \quad (4)$$

where

$$\beta = k_0 v_{gr}^2 |k''| < 1, \quad \gamma = \alpha n_2 |A_0|^2 / 2, \quad \alpha = k_0 z_0 = k_0 T_0 v_{gr}$$

The two non-paraxial terms in the brackets are of the same order. Since equation (4) has two additional terms, to normalize it, it is necessary to take into account two new dimensionless parameters ($\alpha$ and $\beta$). The constants $\alpha$ and $\gamma$ count for the number of oscillations at level *1/e* of the maximum of the amplitude of light pulses and respectively the nonlinearity of the medium. The parameter $\beta$ is connected with dispersion of group velocity. The coefficient ($1/2\alpha$) in front of the expression in brackets is inversely proportional to the initial duration of the pulse $T_0$. Its magnitude, for laser pulses with carrier wavelength $\lambda_0=1,5\mu m$ and different $T_0$, propagating in silica single-mode fibers, is presented in Table 1. For nanosecond and picosecond light pulses this term is quite small and can be neglected. That is way, for these regions NSE describes well the evolution of laser pulses in single-mode fibers. Obviously, for femtosecond optical pulses the expression in brackets must be taken into account.

**Table 1.**

| $T_0$ | $1/2\alpha$ |
|---|---|
| 5 fs | $3,2.10^{-1}$ |
| 10 fs | $1,6.10^{-1}$ |
| 50 fs | $3,2.10^{-2}$ |
| 50 ps | $3,2.10^{-5}$ |
| 50 ns | $3,2.10^{-8}$ |

## 2.2. Solution of the nonlinear scalar amplitude equation

We search for a solution in equation (4) of the kind [20]:

$$U(\tau,\xi) = \Phi(\tau)\exp(ia\xi + ib\tau) \qquad (5)$$

where *a* and *b* are constants. Assuming that the amplitude function $\Phi(\tau)$ is real and replacing the expression (5) in equation (4) the following complex ordinary differential equation is obtained:

$$\frac{|\beta|}{2\alpha}\Phi'' - a\Phi + \frac{ib|\beta|}{\alpha}\Phi' - \frac{b^2|\beta|}{2\alpha}\Phi - \frac{a^2}{2\alpha}\Phi - \frac{ia}{\alpha}\Phi' + \frac{ab}{\alpha}\Phi + \gamma\Phi^3 = 0 \qquad (6)$$

where $\Phi'$ and $\Phi''$ are respectively the first and second derivative of the function $\Phi$ with respect to the variable $\tau$. Our next step is to find the constants *a* and *b*. After equalizing the real and imaginary parts, the following two ordinary differential equations are obtained:

- From the real part:

$$\Phi'' - B\Phi + \frac{2\alpha\gamma}{|\beta|}\Phi^3 = 0 \qquad (7)$$

where

$$B = 2\alpha b - b^2(1-|\beta|) = const \qquad (8)$$

With $\eta = \sqrt{B}$ is presented the amplitude of the soliton and $N^2 = |\beta|/\alpha\gamma$.

- From the imaginary part:

$$\frac{\Phi'}{\alpha}(b|\beta| - a) = 0 \qquad (9)$$

The first derivative $\Phi'$ is nonzero for arbitrary $\tau \neq 0$. From expressions (8) and (9) we can find the constants *a* and *b*:

$$a = b|\beta| \qquad (10)$$

$$b_{1,2} = \frac{\alpha}{1-|\beta|}\left[1 \pm \sqrt{1 - \frac{\eta^2}{\alpha^2}(1-|\beta|)}\right] \qquad (11)$$

The equation (7) has well known soliton solution:

$$\Phi(\tau) = N\eta\,\text{sech}[\eta(\tau - \tau_0)] \qquad (12)$$

Thus, we found that in local time frames GNAE has an exact analytical soliton solution with *sech*-shape and more complicated phase:

$$U(\xi,\tau) = N\eta\,\text{sech}[\eta(\tau - \tau_0)]\exp(i\psi) \qquad (13)$$

where

$$\psi = \frac{\alpha}{1-|\beta|}\left[1 \pm \sqrt{1 - \frac{\eta^2}{\alpha^2}(1-|\beta|)}\right][\xi|\beta| + \tau] \qquad (14)$$

For laser pulses where the conditions $\alpha \gg 1$ and $\beta \ll 1$ are satisfied, the phase of the pulses (14) can be simplified. The square root in (14) can be presented in Taylor series when $(1-|\beta|)/\alpha^2 \ll 1$:

$$1 - \sqrt{1 - \frac{\eta^2}{\alpha^2}(1-|\beta|)} \approx 1 - \left(1 - \frac{\eta^2(1-|\beta|)}{2\alpha^2}\right) = \frac{\eta^2(1-|\beta|)}{2\alpha^2} \qquad (15)$$

For the approximated phase can be used the following expression:

$$\psi \approx \frac{\eta^2}{2\alpha}(\xi|\beta| + \tau) \qquad (16)$$

It is not hard to show that the obtained solution written in Cartesian coordinate system is an exact analytical solution of equation (1) when $\eta=1, N=1$:

$$U(z,t) = \text{sech}\left[\frac{z - v_{gr}t}{z_0}\right]\exp(i\psi) \qquad (17)$$

where

$$\psi = \frac{\alpha}{1-|\beta|}\left[1 \pm \sqrt{1 - \frac{1}{\alpha^2}(1-|\beta|)}\right]\left[\frac{z|\beta|}{z_0} - \frac{z - v_{gr}t}{z_0}\right] \qquad (18)$$

If we use the same approximation that was made for equation (14), we obtained the following expression:

$$\psi = \frac{z|k''|}{2T_0^2} - \frac{1}{2\alpha}\left(\frac{z - v_{gr}t}{z_0}\right) \qquad (19)$$

## 3. Solution of the nonlinear Schrödinger equation

The nonlinear Schrödinger equation describing the propagation of nanosecond and picosecond laser pulses in Kerr-type nonlinear dispersive isotropic medium, presented in local time coordinate system, has the form [1,2,4]:

$$i\frac{\partial A}{\partial z} + \frac{|k''|}{2}\frac{\partial^2 A}{\partial T^2} + k_0 n_2 |A|^2 A = 0 \qquad (20)$$

In order to compare the solutions of GNAE and NSE it is made the same change of variables (3). NSE (20) in dimensionless form is of the kind [1,2,4,20]:

$$i\frac{\partial U}{\partial \xi} + \frac{|\beta|}{2\alpha}\frac{\partial^2 U}{\partial \tau^2} + \gamma |U|^2 U = 0 \qquad (21)$$

The equation (21) has well known soliton solution [4,8]:

$$U(\tau,\xi) = N\eta \operatorname{sech}[\eta(\tau - \tau_0)]\exp(i\psi_{NSE}), \qquad (22)$$

where

$$\psi_{NSE} = \frac{\eta^2 |\beta|\xi}{2\alpha} \qquad (23)$$

Here $\eta$ is the amplitude of the soliton, which does not depend on its velocity. In its physical nature, the soliton propagates as a modulated wave packet in nonlinear dispersive medium with constant velocity without any changes in its spatio-temporal profile for arbitrarily long distances.

The fundamental soliton solution ($\eta=1$, $N=1$) of equation (20) in Cartesian coordinate system can be written in the form:

$$U(t,z) = \operatorname{sech}\left[\frac{z - v_{gr} t}{z_0}\right]\exp\left(i\frac{z|k''|}{2T_0^2}\right) \qquad (24)$$

## 4. Comparison between the non-paraxial solution and NSE one

The modulus of the non-paraxial soliton solution (13) has the same *sech*-shape as that of NSE one. The main difference appears in their phases. The phase of the non-paraxial soliton (14) has a considerably more complicated expression according to the phase of NSE one (23). The non-paraxial solution (14) has an additional term with linear phase modulation of $\tau$. It depends inversely on the number of oscillations at level $1/e$ of the maximum of the amplitude of the pulse. Thus, it can be expected that for higher number of oscillations the additional phase modulation can be neglected and the soliton solution tends to the standard NSE one. On the other hand, this term will be considerable for ultra-short optical pulses with attosecond or femtosecond duration (see Table 1). To show its influence on the evolution of laser pulses, we performed numerical analysis of NSE and GNAE.

In Fig. 1 it is shown the dynamics of a non-paraxial fundamental soliton with 40 oscillations under its envelope (*300 fs*), obtained by solving the GNAE (4). As it can be seen below the evolution of *300 fs* optical pulses does not differ from that of the typical NSE soliton. The shape and the spectrum of the soliton do not change during its propagation.

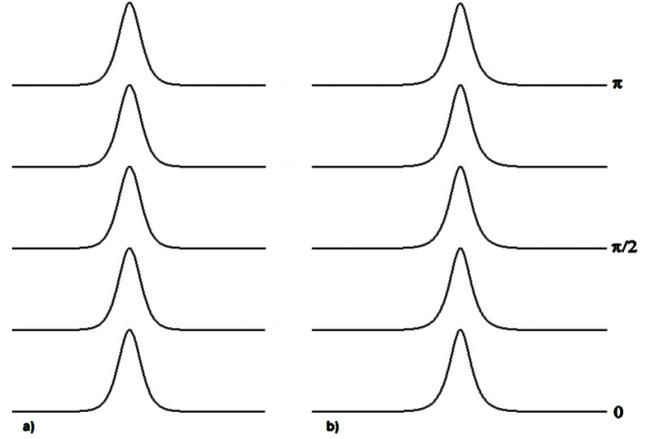

*Figure 1.* Plot of the evolution in $\tau, \xi$ - projection of a) the intensity profile and b) spectrum of 300 fs fundamental non-paraxial soliton (N=1). The numerical results are obtained by solving the nonlinear amplitude equation (4).

In Fig. 2 it is shown the evolution of *300 fs* higher-order soliton (*N=2*) governed again by GNAE (4). From the simulations below we can see that the GNAE (4), used for describing the behavior of narrow-band optical pulses, gives the same results as the NSE (20).

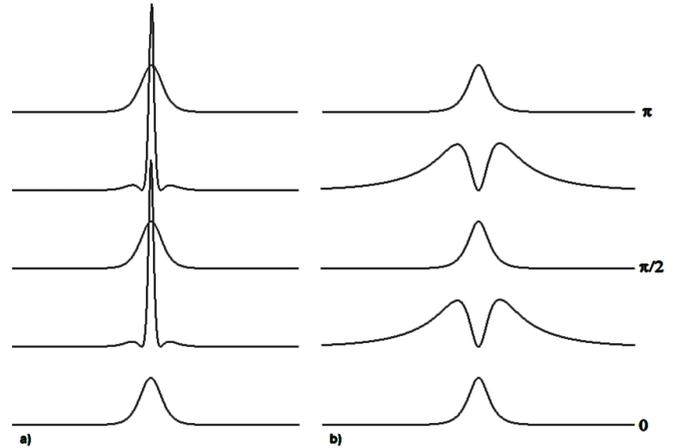

*Figure 2.* Plot of the evolution in $\tau, \xi$ - projection of a) the intensity profile and b) spectrum of 300 fs second-order non-paraxial soliton (N=2). The numerical results are obtained by solving the nonlinear amplitude equation (4). It is clearly seen the typical preciosity that occurs for all higher-order solitons. This is the well-known result observed from NSE approximation.

The situation is quite different in the frames of broad-band optical pulses. In Fig. 3 it is presented the evolution of *14 fs* non-paraxial optical soliton with few oscillations under its envelope. The shape and the spectrum of the pulse do not change during its propagation but we observe a significant temporal shift of the soliton proportional to ($1/2\alpha$). This behavior is a result of the additional phase term found in solution (14). Its influence can be seen only for broad-band pulses.

For higher-order solitons the change in the phase of broad-band laser pulses, propagating in optical fibers, has an essential role. The effect of the additional phase term leads to completely different dynamics in comparison with the two-*sech* NSE solitons, even when the higher order nonlinear and dispersive effects are not included. On Fig. 4 it is presented the evolution of broad-band (*14 fs*) non-paraxial optical soliton of second-order. As a result of the phase shift the soliton breaks up into two components which are following periodical pattern that doesn't match with the evolution of the standard two-soliton solution of NSE with typical periodicity $\pi/2$. The two non-paraxial solitons obtained by solving the GNAE form a bound state. It is not observed a recovery of the initial shape and spectrum of the soliton.

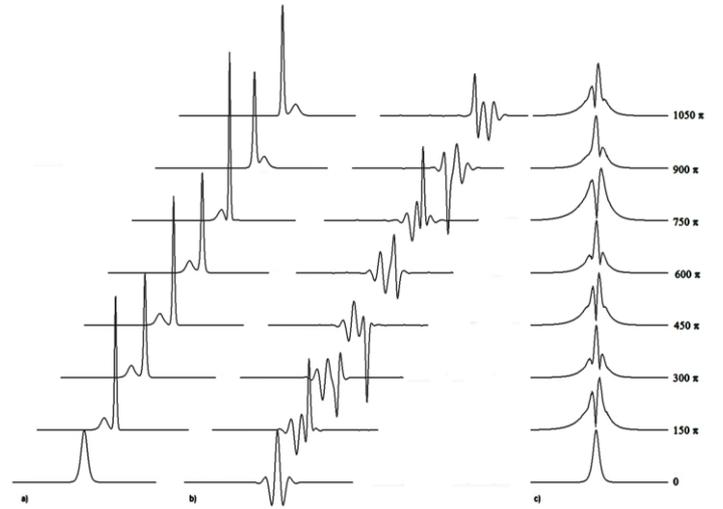

*Figure 4.* Plot of the evolution in $\tau, \xi$ - projection of a) the intensity profile, b) real field normalized to $\pi/2$ and c) the spectrum of 14 fs non-paraxial second-order soliton (N=2). The numerical results are obtained by solving the nonlinear amplitude equation (4).

## 5. Conclusion

In the present work the propagation of optical pulses with narrow and broad-band spectrum in nonlinear regime is investigated in the frames of two different evolution equations: the general nonlinear amplitude equation (4) and the nonlinear Schrödinger equation (21). We found *new* analytical non-paraxial solution of GNAE (4). Its dynamics is compared with that of the well-known NSE. The main difference between the non-paraxial and the standard soliton solution is in their phases. The numerical results lead to the following important conclusions:

- For a large number of oscillations under the envelope (narrow-band case) the non-paraxial soliton obtained by solving GNAE (4) matches very well with the standard soliton solution of NSE (21). That is why NSE is used so widely for describing the behavior of narrow-band light pulses.

- For optical pulses with few oscillations under the envelope (broad-band case) the evolution of the non-paraxial solitons differs from the typical soliton solution of NSE. The shape and the spectrum of a fundamental soliton described by the GNAE do not change during its propagation but a significant temporal shift is observed.

- Higher-order solitons:
    - For a large number of cycles under the envelope, non-paraxial solitons admit the same periodical behavior as that of the two-soliton solution of NSE (see Fig. 2).

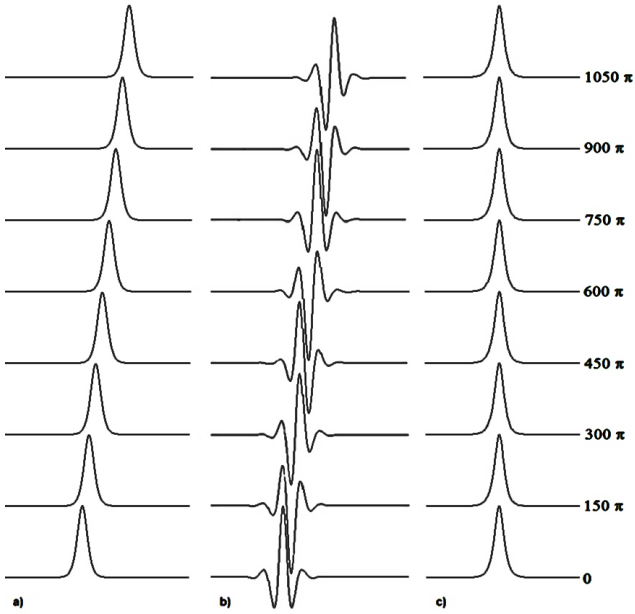

*Figure 3.* Plot of the evolution in $\tau, \xi$ - projection of a) the intensity profile, b) real field, normalized to $\pi/2$ and c) the spectrum of 14 fs non-paraxial fundamental soliton (N=1). The numerical results are obtained by solving the nonlinear amplitude equation (4).

- For a few cycles inside the envelope, the non-paraxial two-soliton solution breaks up into two components which form bound state (see Fig. 4).

In this paper it is shown that GNAE has an exact analytical non-paraxial soliton solution which describes more accurately and correctly the evolution of broad-band optical pulses as well as narrow-band ones in Kerr-type nonlinear dispersive isotropic medium. The obtained results are important for the better understanding and more complete description of the nonlinear dynamics of ultra-short laser pulses propagating in single-mode fibers or planar waveguides.

**Acknowledgments**

The present work is supported by projects NI15FFIT005 Faculty of Physics, Plovdiv University "Paisii Hilendarski", Bulgaria.

*Corresponding author: valerislavchev@yahoo.com